**Wallets as Universal Access Devices**

**Kim Peiter Jørgensen**

*Tecminho, University of Minho,*

*Guimaraes,Portugal*

*kimpeiter@proton.me*

Abstract:
Wallets are access points for the digital economy's value creation. Wallets for blockchains store the end-users' cryptographic keys for administrating their digital assets and enable access to blockchain/ Web3 systems. Web3 delivers new service opportunities. This chapter focuses on the Web3-enabled release of value through the lens of wallets.

Wallets may be implemented as software apps on smartphones, web apps on desktops, or hardware devices. Wallet users request high security, ease of use, and access of relevance from their wallets.

Increasing connectivity, functionality, autonomy, personal support, and off-line capability make the wallet into the user's Universal Access Device for any digital asset. Through wallet-based services, the owner obtains enhanced digital empowerment. The new Web3 solution-areas, Identity and Decentralisation, enable considerable societal effects, and wallets are an integral part of these. One example is self-sovereign identity solutions combined with wallet-borne AI for personalized support, empowering the end-user beyond anything previously known.

Improved welfare is foreseen globally through enlarged markets with collaborative services with drastically lowered transaction costs compared to today, the expected vastly increased levels of automation in society necessitate enhanced end-user protection. As wallets are considered a weak spot for security, improving overall security through blockchains is essential.

Keywords: Wallets for blockchain systems; Recent Wallet developments; Ease-of-Use, Usability; Security; Web3 technology raises several new opportunities; Identity wallets; The decentralization paradigm is fundamental to the original blockchain thinking; Web3 greatly improves interoperability; Trust; Civic solutions where wallets are already contributing to changing and improving human society; The digital divide; Wallets are developing in other roles than traditionally expressed; Future directions of wallet platforms; Economic and societal implications - Who benefits; New business models are necessary; Risks.

## 1.Introduction

Digital wallets are already well-known from Web2 systems like Facebook, Amazon, and Google services. Wallets for blockchain systems are devices, typically dApps software on smartphones, connecting to blockchain systems [Jørgensen & Beck, 2022]. Wallets store the end-user's cryptographic keys for administrating their digital assets and enable access to blockchain/Web3 systems. Even if they resemble the wallets for Web2 systems, wallets for blockchains are fundamentally different. Based on Web3, they enable entirely new service opportunities we can hardly imagine today, combined with a much higher overall security for the end-user [Cai *et al.*, 2018]. The paradigms behind blockchain were early described by Swan, [2019] and [Swan & de Filippi, 2017].

It should be noted that any of the following examples can be constructed outside of Web3 - it is just far more effective, safe, and easy to develop these as Web3 solutions. Decentralization and identity are well-known core abilities enabled by Web3. Other Web3 aspects include drastically improved interoperability [Park *et al*., 2023] and the use of Metaverse technologies.

We will describe the world of wallets through brief narratives to illustrate realized roles and upcoming solutions. We first take a traditional look at wallets for Web3/blockchains as appendages to the blockchain systems, examining how their implementation brings benefits and some limitations of the present solutions. Then, we inspect recent developments with ease of use and security, analyze some of the opportunities raised by the technology with digital identity and decentralized services providing user empowerment, system and societal interconnection, trust services, and consider examples of civic solutions where wallets are already contributing to society. What is coming? The wallet's increased ability to act independently creates entirely new solutions and services with wallets in other roles than those traditionally expressed and utilized.

Finally, we will discuss some of the overall economic and societal implications of Web3 perspectives and summarize our findings in Take-Away conclusions.

**Wallets for blockchain systems.** A traditional look. Popchev *et al*. [2023] define "A blockchain wallet is a mechanism (device, physical medium, software, or a service), operating through cryptographic key pairs, that enables users to interact with a variety of blockchain-based assets and serves as an individual's interface to blockchain networks." Wallets place the user at the center of focus [Voskobojnikov *et al*., 2021]. The user becomes empowered vastly beyond what is known from today's digital wallets.

In short: Wallets are the end-users access points for value creation in the digital economy [Jørgensen & Beck, 2022], as illustrated in the use case for identities by Degen & Teubner, [2024]. Here, the digital service becomes available to what has been called the access economy. The literally thousands of wallets for blockchain systems available cover many usage areas, although the vast majority focus on crypto-currencies, the wallet functionality has expanded drastically in recent years; on top of the crypto-currency solutions are new types of tokens like Non-Fungible Tokens, real-world assets, RWAs, the upcoming identity solutions, and Verifiable Credentials with large-scale use-cases covering everyday life in the European Blockchain Services Infrastructure, EBSI.

Below, a few examples of current wallet functionality and solutions are presented. Through a wallet, users can manage their digital assets,

including their personal key, and manage access by governing which systems are accessed and which data are transferred, with whom and when. These two areas, access, and asset management, can be considered the core wallet functionalities. Several auxiliary services exist to make the end user's lives easier, offering extra storage or enhanced security functionality [Hyperledger Indy, 2024] to wallets accessing their applications.

**Recent Wallet developments.** Further to the above core functions, the wallet fulfills several specific functions dependent on the blockchain system it accesses. With developments within digital identity like self-sovereign Identity (SSI), it is possible for the wallet holder to govern and use the generalized digital identity in completely new service types. Such solutions are being launched on a large scale across the EU with specific Identity Wallets (EUID). [European Commission, 2025].

**Ease-of-Use, Usability.** Ease of use is a priority for the end-users. Usability is so essential for the success of any device or service that it should be taken for granted, together with high security and a well-performing platform for services. Currently, the end-users are insufficiently shielded from the complexities of blockchain technology by not being presented with a simplified, easy-to-use interface [Voskobojnikov *et al.*, 2021]. The need for intuitive use will increase as the types of assets managed through wallets go beyond finance-like identity [Sartor *et al.*, 2022] into other societal areas with verifiable credentials and other digital tokens [Tan *et al.*, 2023]. We need an easy-to-navigate and understand user-centric wallet interface [Hart, 2023], avoiding technical jargon while providing precise and helpful error messages to guide users.

The importance of this is illustrated through the connection between ease of use and reduced digital divide [Loo & Ngan, 2012]. The wallet should work seamlessly across different devices and blockchain systems as part of ease-of-use. Users wish to avoid being bothered by accessing many apps, hence the need for universal wallets [Jørgensen & Beck, 2022]. This should be contrasted with the purpose-specific wallets of today, which work as end-user interfaces to a set of specific applications. In the

Far East, there is a development around super-apps [Minghai *et al.*, 2023], essentially the same analysis we make here for universal wallets where all the functionality requested by the user is made available via one app.

To reach enough end-users and citizens to achieve the intended benefits and usage, authorities and corporations need to plan how to incorporate wallets as the access vector to their digital services [Viriyasitavat *et al.*, 2019]. Albayati *et al*. [2021] find ease of use and the relevance of services, for example, that the wallet adapts to the user's interactions and can provide personalized experiences and security [Yu *et al.*, 2024] to be the highest priorities among the analyzed users.

**Security** is one of the key 'abilities' blockchain systems offer [Goyal, 2023]. Users need to feel safe using their wallets. The wallet is often considered the weak link in very safe blockchain solutions [Houy *et al.*, 2023; Marella *et al.*, 2020] precisely because they have access to the blockchain without being one. It should be noted that good security needs to be closely linked with user-friendliness to avoid end-users making security-compromising errors [Sambin, 2023]. Security risks [Erinle *et al.*, 2023] are quite well understood with a rich literature [Houy *et al.*, 2023]. Further to penetration, a risk is losing the private keys [Erinle *et al.,* 2023], which irreversibly locks the assets governed by that wallet.

Other risks are more subtle. A risk that users should always consider is data framing. For example, anyone can send data segments to any current wallet. These segments might contain incriminating data being planted to discredit the holder.

A key security consideration for Web3 wallets is who controls (owns) the user's keys. Wallet recovery methods are key to user confidence. There are two types of custody for wallets. The first is a custodial wallet, where a third party (often a crypto exchange) manages your private keys. If one loses/forgets one's keys, the access can be recovered. The other type of custody is called non-custodial wallets [Vadlamani & Sharma, 2023]. With a non-custodial wallet, the end-user alone has complete control over the wallet's key and digital access [Perdana & Hu, 2023]. Non-custodial also means there is no safety net for the user. If the keys are lost, there is no way to contact the assets connected to the wallet [Takei & Shudo, 2024]. This becomes relevant for people who may have been but can no longer administer their digital wallets [den Breeijen *et al*., 2022]. Examples of both types of custody could be Metamask (https://metamask.io/), a custodial-type wallet service that also offers a non-custodial wallet. Binance (https://www.binance.com/en/square) is another example of a custodial construct with its Trust wallet as a highly reputed non-custodial open-source offering (https://trustwallet.com/).

The custodial situation is quite different for institutional users which need to be fully backed up in case of inability. Anchorage (https://www.anchorage.com/), for example, offers custodial solutions as a necessity, which also means other security solutions are available for recreating keys if lost. Here, continuous operation is a key criterion rather than a user's privacy.

Wallets may be categorized into three categories according to the medium of their implementation: software, hardware, and paper. The most popular wallets for blockchains are software wallets. They come as solutions on the web, desktop, or mobile. Some software wallets are easy to use, inexpensive or free, and capable of managing a large number of assets. The net connection is an obvious attack vector as they are connected to their assets through the internet as so-called "hot" wallets. Several approaches increase security for software wallets like multi-sig wallets, which require multiple authorizing signatures, adding control and reducing fraudulent transaction risk. An interesting software wallet providing very high levels of security is INTMAX (https://www.intmax.io/), which claims to reach similar levels of protection as hardware wallets [Rybakken et al., 2023] through using homomorphic encryption.

Hardware wallets are physical devices; the private keys they store on a secure device cannot be transferred out of the device in plaintext. They are considered the most secure and are often used for asset storage. Even though crypto-wallet hardware stores keys separately from exchanges, they still interact with Web3 and may inadvertently be used to sign malicious smart contracts that send assets to a hacker. Most hardware wallets are not very user-friendly; they are not free, and most importantly, they introduce a single point of failure: they can be lost, stolen, or damaged. Dabrowski *et al*. [2021] consider "hardware wallets the new single point of failure." Cold wallets are considered the most secure type of crypto wallets because they operate entirely offline and do not have Web3 interaction. However, they can be and are hacked.

The wallet category paper wallets provide off-line ("cold") storage of private keys. Along with the private and public key pair are associated QR codes. In printed form, these enable receiving or spending digital assets. Paper wallets were popular when first introduced, but users soon realized they could be lost, stolen, or damaged.

**Web3 technology raises several new opportunities.** The most significant initial change in system paradigms from these new classes of digital services, not least combined with the proliferation of smart devices [Büttgen *et al*, 2021], will be the introduction of Self Sovereign Identity (SSI) into society [de Amorim, 2024] building upon the Web3 identity tokens, Verifiable Credentials [W3C, 2024] and Decentralized Identifiers DIDs [W3C, 2022]. Unlike the several VC-specialized wallets for the non-blockchain world, these tokens are built for transferability from the outset and do not present challenges for blockchain wallets. The wallet is key for success in this new environment as the user's access to and interaction with these services.

**Identity wallets.** [Podgorelec *et al*., 2022] are a rapidly growing market transforming technological potential into value [World Economic Forum, 2024]. The interest in improving digital identity has been growing for 20 years. A key paper for a new way of thinking about digital identity, placing the user in charge, is by Cameron [2005]. Several countries and organizations have digital identity drives, with the PRC, India, and the EU mentioned here as the currently most significant in volume and impact. The EU is based upon the eIDAS2.0 regulation launching EU-wide identity systems, to which all EU countries need to develop an EUID wallet solution and enable EU citizens to operate their credentials from their country of origin within all other EU countries by 2026

[European Parliament, 2024]. This is spurring a massive number of services with substantial use and interaction with Verifiable Credentials and Decentralized Identifiers [Sedlmeir *et al*., 2021]. These digital identity drives are of key interest to the financial and banking industry, providing much improved Know Your Customer, KYC, and Anti Money Laundering, AML, services. This Web3 proliferation, with identity and decentralization as key elements, enables local decentralized digital services with potentially ultra-low transaction costs as these services oust the expensive current bank and credit card solutions.

**The decentralization paradigm is fundamental to the original blockchain thinking**. With Decentralized Identifiers, DIDs [W3C, 2022], a standard way of referring to an identity is described with a set of transactions conducted locally. DIDs are already used in hospitality and travel [Identity Foundation, 2023], the clothing industry [Guth-Orlowski & Sabadello, 2023], and together with Verifiable Credentials in the upcoming digital product passports in the EU [Garcia *et al.*, 2020].

**Web3 greatly improves interoperability** between IT systems [Park *et al*., 2023; Liu *et al*., 2022] often by interconnecting the blockchains through using of Smart Contracts [Khan *et al.,* 2021]. This approach also let a blockchain govern non-blockchain applications. Popchev *et al.* [2023] mention wallets using Smart Contracts to secure and automate wallet operations. As it can be quite cumbersome to interconnect blockchains directly, using wallets [Schlatt *et al.,* 2023] can simplify cross-blockchain connection, at least at the end-user level. Such solutions demonstrate the possibility of managing more asset types than cryptocurrencies across domain areas like healthcare, smart cities, supply chains, and gaming.

**Trust.** The end users' trust in the digital solutions is crucial to fully joining a digital lifestyle [Marella *et al*., 2020]. Credible decentral structures, including wallets, must be interwoven into the Web3 architecture [Bambacht & Pouwelse, 2022]. Another element in trust is availability. We trust that the services we need are available when needed. With increasing levels of automation, the availability of functionality becomes critical for the functioning of the individual and society. Several solutions for interim off-line capability have been proposed [Igboanusi *et al.*, 2021].

**Civic solutions where wallets are already contributing to changing and improving human society**. An essential aspect of Web3 is an opportunity to spread empowerment of individuals through Self Sovereign Identity and general use of Verifiable Credentials, enabling improved audibility, accountability, and security. These opportunities for participative citizenship are already expressed in several activities where wallets contribute to changing and improving human society. For example, agriculture in Kenya [Bolt, 2019] and Ghana [Glavanits & Szabo, 2024] enable Asare *et al.* [2024] to consider blockchain technology as a catalyst for transformation in Africa. These projects utilize the technical opportunities to address solutions other than those found before Web3, with the wallet playing a key role precisely where the end-user is involved. Such civic solutions are initiated across the world with wallets constructed to handle tokens created for use locally by the local community [Balbo *et al.*, 2020].

Decentralisation. A key point for understanding the wallet perspective in civic or democratic blockchain projects is blockchain's fundamentally decentralized approach. We have seen how wallets with blockchains enable the user to safely participate in far more digital activities than today. "Disciplined initiative" [Alexander, 2020] states that the person in the situation has the most precise perspective of what is needed and what is doable within the plans and resources available. Blockchain wallets' user-centric and empowering approach also allows for transactional optimization, which is a significant part of decentralized systems' overall much-lowered transaction cost. Examples of socioeconomic part-taking and interactions are reported from local communities [Viano *et al.*, 2022] and smart cities [Kassen, 2021]. Social coins, e.g., tokens with some value attached, are found as loyalty tokens in various shopping from gaming and social gatherings [Abbas & Alkhzaimi, 2024], who also report potential interchange with fiat money. Viano *et al*. [2023] provide an example of community wallets with a tokens-based decentralized economy.

This local societal participation can be an element in the democratic process, including voting [Ghazinoory *et al*., 2024], and opens new ways of accessing and participating in society. The massive number of blockchain solutions in the community area in China should also be observed [Kshetri, 2023].

Inclusion of marginal groups. A further example to the community solutions mentioned above are the many solutions for refugees in refugee camps [Garazha, 2024].

**The digital divide,** the growing split between those who can and do use the digital systems available and those outside these categories, has become more critical with the increasing levels of automation. As wallets cover an increasing amount of each person's digital life activities, it becomes critical to reduce the digital divide actively. One reason for this split is unequal access to digital technology, including smartphones and devices needed to connect to the services. Another is that society is changing with the many new systems being implemented. A person cannot participate anymore without being connected to the societal systems [Helsper, 2021]. The reservations of many people not currently using systems are often related to security and privacy; one could say they lack trust in the system, hence the importance of the Self Sovereignty concept and generally the security and protection of users' assets and privacy in a decentralized environment.

One of the key activities needed to bridge this gap is massive education to prepare and train the population for these new societal practices [Vassilakopoulou & Hustad, 2023]. It is not sufficient just to provide instructions on how to use the latest systems; the concept of using a wallet as an access and management tool to participate as a digital citizen is new to a high percentage of the population. This is demonstrated in the Portuguese Blockchain Agenda [2022] where a nationwide blockchain-based set of societal solutions is developed. Here is a clear need for a strong focus on end-user training to make people use the wallets and new services efficiently and securely. Massive online training seems the only realistic way to achieve this.

Access for impaired people. In a highly automated society where all users become highly dependent on their wallets, a key question is how to integrate users who cannot use these facilities. Previously, we have pointed to custodial solutions for people who, for a reason, are mentally unable to use their wallets and the relevant system. Physical impairment also prevents citizens from using the systems. Of the many areas of impairments and practical solutions to mitigate these, we mention one: visual impairment [Zhou *et al.*, 2023].

Anti-crime activities already being realized, with quite successful anti-corruption drives are reported from the public sector [Olalekan, 2024] and accounting and auditing from the private sector [Kaplan, 2021] as well as successful anti-trafficking activities [Enrile & Aquino-Adriatico, 2024].

What is coming? Wallets are also becoming the wallet-carrier's process coordinator, where smart contracts may provide an easy way to coordinate processes among solutions on different IT platforms as an interconnectivity device. The wallet is not a blockchain; this can be exploited for enhanced interconnection to legacy and non-blockchain systems. That is needed as most systems today accessed by end-users are still non-blockchain systems further to the blockchain systems already in place. Such interconnectivity can be developed into an actual collaboration between several parties exploiting the abilities of Web3 solutions. This is key for local and rural development of community solutions.

**Wallets are developing in other roles than traditionally expressed** with drastically increasing levels of automation and the necessarily associated automation of security procedures such as sign-on and continuous validation, the platform also enables far more personal support from decentralized AI directly on the wallet platform. Kin is an early example of this (https://mykin.ai/), and Apple is expected to follow suit, establishing customer bots [Birch, 2023]. This demonstrates the proliferation of areas where wallets will cover increasing aspects of digital service and support for the end-user. Are wallets becoming agents? Barresi & Zatti [2020] point to the logical development of the wallet as the Universal Access Device for financial services. Generalizing this, Büttgen *et al.* [2021] point to the need for autonomous, proactive wallets to access the drastically increasing number of digital services each person will meet in the future with increased levels of services using smart devices. The universal wallet is a key access and digital asset mechanism in all these examples. The Web3 technology stack represents the next leap forward in institutional technology: Crypto technologies facilitate trust and lower the costs of economic coordination. This innovation enables elements to "talk" to each other as the wallets may cooperate in swarms, communicating wallet-to-wallet. Notice that these multi-agent systems could behave as decentralized artificial superintelligence even without the individual elements of the network necessarily being intelligent [Ponomarev & Voronkov, 2017], so even before a general spread of AI to wallets, the system could behave as intelligent [Shammar *et al.*, 2024]. Such a solution has already been launched to enhance security [Strobel *et al.*, 2023].

The wallet will be developed and refined with the progress of society and as more systems are re-thought as decentralized systems to obtain the drastically lowered transaction and intermediation costs [Chen & Bellavitis, 2020] and enhanced types of services made possible through Web3-enabled interaction.

**Future directions of wallet platforms.** Smartphone-based wallets are by far the most prevalent and commonly used today. Their functionality could be ported to platforms that are more convenient for use, like Heads Up Computing. The number of wearable devices is rapidly expanding [Ometov *et al.*, 2021], differentiating from smartphones [Zhao *et al.*, 2023], offering seamless computing support for humans' daily activities as the next milestone. The wallet and many other devices are logical devices delivering specific functionality for access- and asset management and some potentially collaborative interconnectivity. Such new processing platforms fit into current IoT, and smart devices are soon to be superseded by really smart devices that are expected to increase in number in the coming years. These smart devices will need faculty to decide what to access and how to administer the relevant assets within their realm. For this, they will need wallet functionality. Such embedded wallets already exist [Watanabe *et al.*, 2024]. With autonomous and proactive solutions [Santana & Albareda, 2022], the wallet may be acting on its creation of entirely new solutions and services with functionality previously considered unthinkable.

The above wallet functionalities are conceptualized in the figure below.

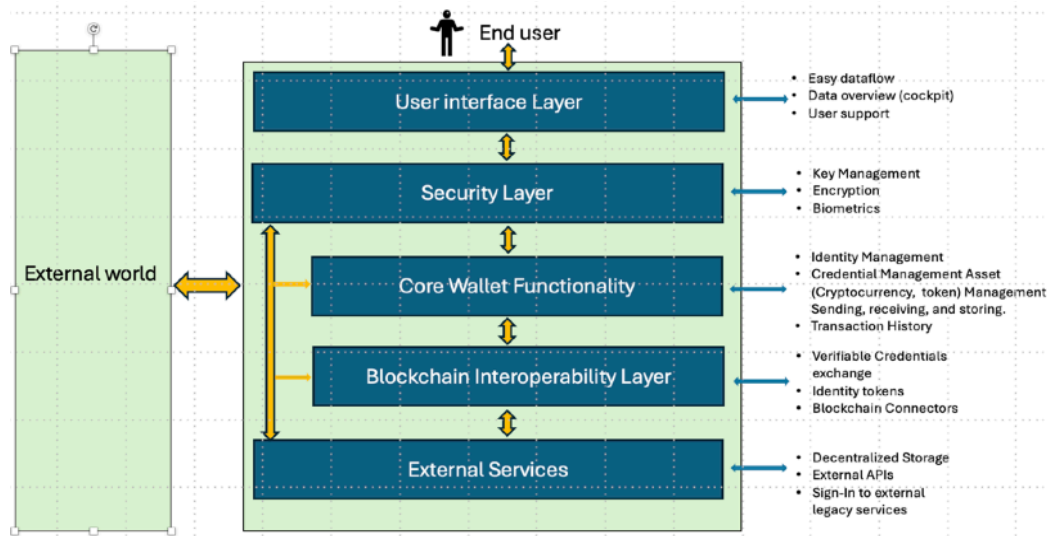

Figure: Overview Blueprint for EndUser Wallet

Economic and societal implications - Who benefits? Blockchain wallets represent several interesting aspects of delivering the benefits presented here. The benefits from a transactional economic perspective are not difficult to estimate. Wallet's initial role as governance and operating devices around cryptocurrencies, with the spread of crypto assets into general tokens, such as Non-Fungible Tokens, NFTs, and Real-World

Assets, RWAs, are examples fitting into classical business models where value build-up is identifiable.

Web3 services such as Digital Identity and particularly Self Sovereign Identity empower the users to maintain control over the information associated with their identity, increasing security and reducing data leakage. With increased functionality and security, improved ease of use, reduced transaction costs, and a much-increased number of new services and new types of digital assets to manage, the end-user can enter the digital world as self-sovereign with the wallet as the direct interface to a vast number of new services. The wallet carrying digital identity files may provide far more streamlined and secure digital services than those known today. With the proliferation of Identity Wallets - EUID, for example - some of the described benefits are classical, like slim-lined payments as well as direct benefits from improved KYC and AML with an expected reduction in data theft and other fraud areas amounting to annual savings of trillions of $s and €s. In all these cases, there are direct transactional benefits. Further, some are like insurance, where you do not directly see the benefit if the mechanism works or if theft or fraud is avoided.

It is necessary to look into the nature of value projection, e.g. who benefits? Do the benefits materialize at the microeconomic, ecosystem, meso-, or at the macro-level? Wallets are of direct importance as user-access points for the individual systems (microeconomic consideration) and at the economic ecosystem level since the systems are interconnected. Expanding its use into new industries will provide network effects at a meso-level. Finally, there is a much larger national/ global context (macroeconomic consideration) as including a more significant percentage of the population as wallet users will enhance the productivity of society as a whole. The wallet, as an essential element being a site of value creation in the rapidly growing software-based access economy, is rarely considered in the above scenarios. Value creation in wallets should be seen in the context of wallets being connectable to many services, not just the marginal value of each wallet connection summed up over all the systems.

Having a mass of active wallet users provides a better digitally trained population and cumulative benefits from digitization. A parallel is the welfare state, where the certainty of support makes life easier, more predictable, and less expensive for the individual user to the benefit of the individual, the local community, and the nation. The actual beneficiaries of the value creation of wallets are thus partly society, the end-user, and the digital services involved. This argumentation is key in this chapter; a well-designed wallet could benefit everyone in society, though in different ways.

A list of benefits for key high-level groups, a could be:

Individual citizens: Greater security, easier access to services, and increased financial and social inclusion benefit individuals across all socioeconomic groups. Blockchain wallets can provide access to financial services for those currently excluded from the traditional banking system, empowering them to participate more fully in the economy [Rosli *et al.*, 2022].

Businesses: New markets and opportunities could emerge soon, initially in the technology and finance sectors, but increasingly in other areas of society covering/developing new opportunities enabling collaboration, security, ease of use,

Governments: More efficient service delivery, reduced fraud, and improved citizen satisfaction are potential benefits for governments.

Society as a whole: A long-term outcome could be a more inclusive, secure, and efficient society with reduced inequality. While blockchain technology is safe, widespread adoption will require robust safeguards against new threats that may emerge. As it is almost impossible to tamper with blockchain data, a high degree of trust is provided. There is value in data being resilient in society. The risk of surveillance, whether conducted by private or governmental bodies, is a key concern for users' trust [Torres *et al.*, 2023].

Societal benefits: Reduced data theft: Enhanced security could protect sensitive personal and financial information, benefiting all citizens but particularly vulnerable groups often being targets of fraud and exploitation, reducing social tension.

Increased access to services: Easier access to online services, such as government assistance programs, could streamline processes and reduce bureaucratic hurdles for those needing them most. Custodial and non-custodial wallets were previously treated. The key questions are: How does society provide the necessary support in the social system for recovery mechanisms, custodial care, and general support for the public in maintaining their data over their life cycle?

Automation and efficiency: Blockchain technology can automate processes, facilitate inter-organizational automation, reduce administrative costs.

Challenges. Digital literacy and access are prerequisites for success. Ensuring everyone has the skills needed to use blockchain wallets is crucial. Mass education is necessary to mitigate this divide. As presented here, the wallet concept can be regarded as a societal infrastructure, far more than individual blockchains. It also pays to optimize for the user rather than just for the specific blockchain system, as is standard today.

It is also evident that value generated by wallets shifts significantly if one takes the wallet's end-user perspective from creating value in the specific ecosystem of the requested digital service to the multiplicative effects of values from many elements, as we find in network economic models. An important societal point is often missed: An overall stable society with a high number of services available is generally wealth-generating and beneficial. Think of Pax Romana.

In regionwide implementations, as are currently conducted in the EU, the wallet is key to success "in this new tech/social/regulatory disruptive environment" [de Amorim, 2024], as the user's access to and interaction with these services. Large tech companies have successfully launched wallets that access their web2 applications and have significant market shares that federated identity structures have already achieved. Apple is negotiating with the EU about launching their well-functioning non-eIDAS compliant Digital Driver's License solution in the EU. Many politicians (Personal communication with MEP) and decision-makers understand why eIDAS compliance is important. A benefits realization model seems much needed here to evaluate the benefits' significance and understand the consequences of the various flavors of proposals.

The EU's launch of identity solutions and a large set of verifiable credential-based solutions is an interesting innovation regarding societal value generation. The concept of using a wallet as an access and management tool to participate as a digital citizen for a high percentage of the population is new.

**New business models are necessary** to enhance work on developing decentral business models. Currently, the focus is on local decentralized digital services with ultra-low transaction costs. However, we need to understand and calculate the benefits, even if they spread at many levels in society and across industries and nations. How do we develop business models also with a collaborative aspect for these nation- and international-wide benefit-generating systems, such as wallet services, that can interest the average institutional investor? An interesting aspect is the many well-functioning blockchain solutions provided without a direct profit/loss perspective. This un-monetization could explain the difficulty of financing such solutions even as they generate net value. There is a need for better business models that incorporate these new functionalities and demonstrate to investors where realizable benefits are generated. One example is local functionality, like wallet-borne personal AI. This recent and exciting area is in the early stages of development for local implementation. It will be possible to offer highly personalized and context-related just-in-time support and education programs to the end user's wallet. Even before AI functionality becomes fully available locally, it can be offered as web-based services or local functionality using traditional tools. These will further empower the end-user and open for entirely new services and markets.

Trust in the overall digital solutions is critical for the end user's acceptance of the solutions to make them fully join a digital lifestyle.

The wallet itself can significantly minimize the digital footprint - but it cannot verify if data sent from or via the wallet has been eavesdropped. When you are requested at a border to download everything from your laptop or mobile, you'll just do it - or not enter.

**Risks.** are present as with all systems. One generated by the nationwide perspective promoted here is the need for continuous availability of these systems, which are becoming critical national infrastructure. The blockchain architecture is probably our closest current architecture for continuously available services. Similar performance should be demanded for all service vendors offering nation-critical services.

Also, there is an increasing need for off-line functionality. With the increasing digital service density in society, these services must be available anytime and anywhere. The current blockchain wallet solutions presume increasingly available network services, although shorter or longer off-line situations are relevant for many application areas. Examples illustrating the breadth of these cases are, healthcare, homecare, constructing sites, military), expeditions, and refugees [Garazha, 2024].

New services to assist and empower the end-user will be created, triggering completely new industries from possibilities like vastly enhanced storage, such as IPFS the interplanetary file system. This is needed with the much larger responsibilities placed upon each user. With the new expectations that we all take charge of our total dataset over our lifetime, many practical burdens are thrown upon the end user. How does one find a specific data item from when one was 12 years old? What to do in case of doubt? A call to a support person will typically be too expensive. Automated services must be developed to provide the end-user with the services known from data centers for tracking and restoring problematic data.

Conclusion. The role of wallets in the blockchain/Web3 ecosystem is crucial and often underestimated or not properly understood, particularly when it comes to precise value-creating estimates. As wallets are the gateway to the end-user's access and interaction with the new digital world, they enable the blockchain solutions focusing on end-users to reach their market [Rifkin, 2002], allowing for an access economy. Further, as the wallets may be enabled to access the digital services currently embalmed in legacy systems, their value increases even more. The universal blockchain wallets transform how we interact with the digital world, allowing the user to plunge deeper into a digital lifestyle that is self-governed and self-sovereign. By prioritizing usability and accessibility, we can unlock the power of this technology for everyone. Of course, there are challenges and risks of Web3. However, the overall opinion is that Web3 technology provides orders of magnitude higher security and safety than what is seen on the market today. The dangers are in society as a whole, from drastically increased levels of automation and dependency upon our systems. We have deliberately focused on concrete examples of benefits as they are the drivers for venturing into new projects, and since Web3 solutions often cover new areas and often from other angles and opportunities for connecting to everything rather than the isolated silo approaches of traditional IT solutions. The economic considerations are somewhat qualitative and indicative as our current understanding of some of the most significant potentials is just dawning. We relied heavily on inference in the discussion. That is necessary when going beyond the current frontline - a firmer deductive discussion may come when our models have covered the new grounds. Thus, we have here attempted to establish the anatomy of value generation in decentral networked systems.

Of course, wallets are not everything. However, hopefully, the above chapter has demonstrated that blockchain wallets are the end-users access point to digital services in the Web3 world and the key to releasing the value from Web3/blockchain systems and connecting the end-user to the digital world, blockchain, Web3, and legacy systems. These possibilities open several quite interesting paths to new areas. All this makes it worthwhile to consider the wallet as a sound foundation for creating and utilizing digital services now and in the future.

## References

Abbas, C. and Alkhzaimi, H., 2024. Transforming Economies and Societies: The Potential of Social Coins. *Transforming Economies and Societies: The Potential of Social Coins (November 16, 2024)*.ssrn-5022944. Accessed 12. August 2024.

Albayati, H., Kim, S.K. and Rho, J.J., 2021. A study on the use of cryptocurrency wallets from a user experiences perspective. *Human Behavior and Emerging Technologies, 3*(5), pp.720-738.

Alexander, M.R.P., 2020 *Mission Command: The Need for Disciplined Initiative*. School of Advanced Military Studies US Army Command and General Staff College.

Asare, M.T., Damoah, D.D., Doe, M., Amponsah, R.A., Okoe, E.A. and Amoako, P.Y.O., 2024, September. Blockchain Technology: A Catalyst for Transformation in Africa. In *2024 IEEE SmartBlock4Africa* (pp. 1-5). IEEE.

Balbo, S., Boella, G., Busacchi, P., Cordero, A., De Carne, L., Di Caro, D., Guffanti, A., Mioli, M., Sanino, A. and Schifanella, C., 2020, August. CommonsHood: A Blockchain-based wallet app for local communities. In *2020 IEEE International Conference on Decentralized Applications and Infrastructures (DAPPS)* (pp. 139-144). IEEE.

Bambacht, J. and Pouwelse, J., 2022. Web3: A decentralized societal infrastructure for identity, trust, money, and data. *arXiv preprint arXiv:2203.00398*.

Barresi, R.G. and Zatti, F., 2020. The importance of where central bank digital currencies are custodied: Exploring the need of a universal access device. *Available at SSRN 3691263*.

Birch, D (2023). Identity, Innovation And Very Smart Wallets. Available at https://www.forbes.com/sites/davidbirch/2023/10/11/buterin-innovation-and-very-smart-wallets/. Accessed 12. August 2024.

Bolt, J.S., 2019. *Financial resilience of Kenyan smallholders affected by climate change, and the potential for blockchain technology*. CCAFS.

Büttgen, M., Dicenta, J., Spohrer, K., Venkatesh, V., Raman, R., Hoehle, H., De Keyser, A., Verbeeck, C., Zwienenberg, T.J., Jørgensen, K.P. and Beck, R., 2021. Blockchain in service management and service research–developing a research agenda and managerial implications. *Journal of Service Management Research*, 5(2), pp.71-102.

Cai, W., Wang, Z., Ernst, J.B., Hong, Z., Feng, C. and Leung, V.C., 2018. Decentralized applications: The blockchain-empowered software system. *IEEE access*, 6, pp.53019-53033.

Cameron, K., 2005. The laws of identity. *Microsoft Corp*, 12, pp.8-11.

Chen, Y. and Bellavitis, C., 2020. Blockchain disruption and decentralized finance: The rise of decentralized business models. *Journal of Business Venturing Insights*, 13, p.e00151.

Dabrowski, A., Pfeffer, K., Reichel, M., Mai, A., Weippl, E.R. and Franz, M., 2021, November. Better keep cash in your boots-hardware wallets are the new single point of failure. In *Proceedings of the 2021 ACM CCS Workshop on Decentralized Finance and Security* (pp. 1-8).

de Amorim, A.P. and Sousa, R.D., 2024. The Neutrality and Dichotomy of Self-Sovereign Identity: An Exploratory Study. aisel.aisnet.org/mcis2024/50/.

Degen, K. and Teubner, T., 2024. Wallet wars or digital public infrastructure? Orchestrating a digital identity data ecosystem from a government perspective. *Electronic Markets*, 34(1), p.50.

Den Breeijen, S., van Dijck, G., Jonkers, T., Joosten, R. and Zimmermann, K., 2022. Self-Sovereign Identity and Guardianship in Practice. European Journal of Law and Technology, 13(3).

Enrile, A. and Aquino-Adriatico, G., 2024. Technology Innovations in Fighting Slavery and Human Trafficking. In *The Palgrave Handbook on Modern Slavery* (pp. 179-203). Palgrave Macmillan, Cham.

Erinle, Y., Kethepalli, Y., Feng, Y. and Xu, J., 2023. SoK: Design, vulnerabilities and defense of cryptocurrency wallets. arXiv preprint *arXiv:2307.12874*.

European Commission (2025). https://ec.europa.eu/digital-building-blocks/sites/display/EUDIGITALIDENTITYWALLET/EU+Digital+Identity+Wallet+Home accessed Thursday, 9 January 2025.

European Parliament (2024) Revision of the eIDAS Regulation – European Digital Identity (EUid). In "A Europe Fit for the Digital Age". Available at https://www.europarl.europa.eu/legislative-train/spotlight-JD22/file-eid. Accessed 12. August 2024.

Garazha, A., Merz, C., Schwabe, G. and Zavolokina, L., 2024. Resilience in Times of Crisis: Empowering Refugees with Self-Sovereign Identity.

García, I., Muñoz-Escoí, F. D., Arjona Aroca, J., & Fernández-Bravo Peñuela, F. J. (2024). Digital Product Passport Management with Decentralised Identifiers and Verifiable Credentials. arXiv e-prints, arXiv-2410.

Ghazinoory, S., Mardani, A., Maddah-Ali, M. A., & Montazer, G. A. (2024). A blockchain-powered e-cognocracy model for democratic decision making. Information Systems and e-Business Management, 1-38.

Glavanits, J. and Szabo, T., 2024. FinTech Solutions Supporting Sustainable Agriculture? Lessons from Africa. *Proceedings of the Central and Eastern European eDem and eGov Days 2024*, pp.97-103.

Goyal, A., 2023. Blockchain Wallet for Secure Transactions. *Kilby*, *100*, p.7th. ssrn-4487894.

Guth-Orlowski,Susanne and Sabadello, Markus 2023) https://medium.com/@susi.guth/resolving-digital-product-passports-f08ac8f11b5a. Accessed 11. October 2024.

Hart, S.U., 2024. From promise to practice: A cross-institutional analysis of design trends, enablers and challenges in blockchain-enabled cash and voucher delivery. International Journal of Disaster Risk Reduction, 100, p.104129.

Helsper, E., 2021. The digital disconnect: The social causes and consequences of digital inequalities.

Houy, S., Schmid, P. and Bartel, A., 2023. Security aspects of cryptocurrency wallets—a systematic literature review. *ACM Computing Surveys*, *56*(1), pp.1-31.

Hyperledger Indy (2024). https://hyperledger-indy.readthedocs.io/projects/sdk/en/latest/docs/design/003-wallet-storage/README.html accessed Thursday, 9 January 2025.

Identity Foundation (2023). https://identity.foundation/Hospitality-and-Travel-SIG/spec/. Accessed 19. September 2024.

Igboanusi, I.S., Dirgantoro, K.P., Lee, J.M. and Kim, D.S., 2021. Blockchain side implementation of pure wallet (pw): An offline transaction architecture. *ICT Express*, *7*(3), pp.327-334.

Jørgensen, K.P. and Beck, R., 2022. Universal wallets. Business & Information Systems Engineering, pp.1-11.

Kaplan, A., 2021. Cryptocurrency and corruption: Auditing with blockchain. In *Auditing Ecosystem and Strategic Accounting in the Digital Era: Global Approaches and New Opportunities* (pp. 325-338). Cham: Springer International Publishing.

Kassen, M., 2021. Understanding decentralized civic engagement: Focus on peer-to-peer and blockchain-driven perspectives on e-participation. *Technology in Society*, *66*, p.101650.

Khan, S., Amin, M.B., Azar, A.T. and Aslam, S., 2021. Towards interoperable blockchains: A survey on the role of smart contracts in blockchain interoperability. *IEEE Access*, *9*, pp.116672-116691.

Kshetri, N., 2023. Blockchains with Chinese characteristics. In *Blockchain in the Global South: Opportunities and Challenges for Businesses and Societies* (pp. 83-111). Cham: Springer Nature Switzerland.

Liu, Z., Xiang, Y., Shi, J., Gao, P., Wang, H., Xiao, X., Wen, B., Li, Q. and Hu, Y.C., 2021. Make web3. 0 connected. *IEEE transactions on dependable and secure computing*, *19*(5), pp.2965-2981.

Loo, B.P. and Ngan, Y.L., 2012. Developing mobile telecommunications to narrow digital divide in developing countries? Some lessons from China. *Telecommunications Policy*, *36*(10-11), pp.888-900.

Marella, V., Upreti, B., Merikivi, J. and Tuunainen, V.K., 2020. Understanding the creation of trust in cryptocurrencies: The case of Bitcoin. *Electronic Markets*, *30*(2), pp.259-271.

Minghai, Y., Wenqing, L., Khan, W.A. and Nurhalim, W., 2023. The SuperApp Implementation in Business: Revolutionizing Business Operations for a Seamless Future. *Bincang Sains dan Teknologi*, *2*(03), pp.118-123.

Olalekan, O.A., 2024. BLOCKCHAIN TECHNOLOGY AND ANTI-CORRUPTION MEASURES IN THE SETTING OF PUBLIC ADMINISTRATION IN NIGERIA.

Ometov, A., Chukhno, O., Chukhno, N., Nurmi, J. and Lohan, E.S., 2021, May. When wearable technology meets computing in future networks: A road ahead. In *Proceedings of the 18th ACM International Conference on Computing Frontiers* (pp. 185-190).

Park, A., Wilson, M., Robson, K., Demetis, D. and Kietzmann, J., 2023. Interoperability: Our exciting and terrifying Web3 future. *Business Horizons*, *66*(4), pp.529-541.

Perdana, A. and HU, E.I., 2023. Decentralized Finance (DeFi), Strengths Become Weaknesses: a Literature Survey. *Jurnal RESTI (Rekayasa Sistem dan Teknologi Informasi)*, *7*(2), pp.397-404.

Podgorelec, B., Alber, L. and Zefferer, T., 2022, June. What is a (digital) identity wallet? a systematic literature review. In *2022 IEEE 46th Annual Computers, Software, and Applications Conference (COMPSAC)* (pp. 809-818). IEEE.

Ponomarev, S. and Voronkov, A.E., 2017. Multi-agent systems and decentralized artificial superintelligence. *arXiv preprint arXiv:1702.08529*.

Popchev, I., Radeva, I. and Dimitrova, M., 2023, October. Towards blockchain wallets classification and implementation. In *2023 International Conference Automatics and Informatics (ICAI)* (pp. 346-351). IEEE.

Portuguese Blockchain Agenda 2022. https://www.inesc-id.pt/blockchainpt/ Accessed 12 January 2025.

Rifkin, J., 2001. The age of access: The new culture of hypercapitalism. *Where all of life is a paid for experience/JP Tarcher*.

Rosli, M.S., Saleh, N.S., Md. Ali, A. and Abu Bakar, S., 2023. Factors determining the acceptance of E-wallet among gen Z from the lens of the extended technology acceptance model. *Sustainability*, *15*(7), p.5752.

Rybakken, E., Hioki, L. and Yaksetig, M., 2023. Intmax2: A ZK-rollup with Minimal Onchain Data and Computation Costs Featuring Decentralized Aggregators. *Cryptology ePrint Archive*.

Sambin, G., 2023. *Usability of Safety Critical Applications in Enterprise Environments: Defining Guidelines for Error Preventing UI/UX Patterns and Improving Existing Interfaces* (Doctoral dissertation, Politecnico di Torino).

Santana, C. and Albareda, L., 2022. Blockchain and the emergence of Decentralized Autonomous Organizations (DAOs): An integrative model and research agenda. *Technological Forecasting and Social Change*, *182*, p.121806.

Sartor, S., Sedlmeir, J., Rieger, A. and Roth, T., 2022, June. Love at First Sight? A User Experience Study of Self-Sovereign Identity Wallets. In *ECIS 2022*.

Schlatt, V., Sedlmeir, J., Traue, J. and Völter, F., 2023. Harmonizing sensitive data exchange and double-spending prevention through blockchain and digital wallets: The case of e-prescription management. *Distributed Ledger Technologies: Research and Practice*, *2*(1), pp.1-31.

Sedlmeir, J., Smethurst, R., Rieger, A. and Fridgen, G., 2021. Digital identities and verifiable credentials. *Business & Information Systems Engineering*, *63*(5), pp.603-613.

Shammar, E., Cui, X. and Al-qaness, M.A., 2024. Swarm Learning: A Survey of Concepts, Applications, and Trends. *arXiv preprint arXiv:2405.00556*.

Strobel, V., Pacheco, A. and Dorigo, M., 2023. Robot swarms neutralize harmful Byzantine robots using a blockchain-based token economy. *Science Robotics*, *8*(79), p.eabm4636.

Swan, M., 2019. Blockchain economic networks: Economic network theory —Systemic risk and blockchain technology. *Business Transformation through Blockchain: Volume I*, pp.3-45.

Swan, M. and De Filippi, P., 2017. Towards a philosophy of blockchain. *Metaphilosophy*, *48*.

Takei, Y. and Shudo, K., 2024. Pragmatic Analysis of Key Management for Cryptocurrency Custodians.

Tan, E., Lerouge, E., Du Caju, J. and Du Seuil, D., 2023. Verification of education credentials on European Blockchain Services Infrastructure (EBSI): action research in a cross-border use case between Belgium and Italy. *Big Data and Cognitive Computing*, *7*(2), p.79.

Torres, C.F., Willi, F. and Shinde, S., 2023. Is your wallet snitching on you? an analysis on the privacy implications of web3. In *32nd USENIX Security Symposium (USENIX Security 23)* (pp. 769-786).

Vadlamani, A. and Sharma, S., 2023. Bridging the Divide between DeFi and Regulators: Showcasing Decentralized Autonomous Governance as the Future for Self-Custody Wallet Regulation. *U. Ill. JL Tech. & Pol'y*, p.373.

Vassilakopoulou, P. and Hustad, E., 2023. Bridging digital divides: A literature review and research agenda for information systems research. *Information Systems Frontiers*, *25*(3), pp.955-969.

Viano, C., Avanzo, S., Cerutti, M., Cordero, A., Schifanella, C. and Boella, G., 2022. Blockchain tools for socio-economic interactions in local communities. *Policy and Society*, *41*(3), pp.373-385.

Viano, C., Avanzo, S., Boella, G., Schifanella, C. and Giorgino, V., 2023. Civic Blockchain: Making blockchains accessible for social collaborative economies. *Journal of Responsible Technology*, *15*, p.100066.

Viriyasitavat, W., Da Xu, L., Bi, Z. and Pungpapong, V., 2019. Blockchain and internet of things for modern business process in digital economy—the state of the art. *IEEE transactions on computational social systems*, *6*(6), pp.1420-1432.

Voskobojnikov, A., Wiese, O., Mehrabi Koushki, M., Roth, V. and Beznosov, K., 2021, May. The u in crypto stands for usable: An empirical study of user experience with mobile cryptocurrency wallets. In *Proceedings of the 2021 CHI Conference on Human Factors in Computing Systems* (pp. 1-14).

W3C. (2024). Verifiable Credentials Overview. W3C Group Note 06 July 2024. Available at https://www.w3.org/TR/2024/NOTE-vc-overview-20240706/. Accessed 12. August 2024.

W3C. (2022). Decentralized Identifiers (DIDs) v1.0 Core architecture, data model, and representations. W3C Recommendation 19 July 2022. Available at https://www.w3.org/TR/2022/REC-did-core-20220719/. Accessed 12. August 2024.

Watanabe, H., Ichihara, K. and Aita, T., 2024. VELLET: Verifiable Embedded Wallet for Securing Authenticity and Integrity. *arXiv preprint arXiv:2404.03874.*

World Economic Forum. 2024.How universal wallets can help businesses unleash the full potential of digital identity. Available at https://www.weforum.org/agenda/2024/02/how-businesses-could-use-universal-wallets-to-unleash-potential-of-digital-identity/. Accessed 15 July, 2024.

Yu, Y., Sharma, T., Das, S. and Wang, Y., 2024, May. " Don't put all your eggs in one basket": How Cryptocurrency Users Choose and Secure Their Wallets. In Proceedings of the CHI Conference on Human Factors in Computing Systems (pp. 1-17).

Zhao, S., Tan, F. and Fennedy, K., 2023. Heads-Up Computing Moving Beyond the Device-Centered Paradigm. Communications of the ACM, 66(9), pp.56-63.

Zhou, Z., Sharma, T., Emano, L., Das, S. and Wang, Y., 2023. Iterative design of an accessible crypto wallet for blind users. In Nineteenth Symposium on Usable Privacy and Security (SOUPS 2023) (pp. 381-398).